\documentstyle[prl,aps,multicol]{revtex}
\def\subequations{\begin{mathletters}}
\def\endsubequations{\end{mathletters}}
\begin{document}
\draft
\title{Mean Magnetic Field and Noise Cross-Correlation in Magnetohydrodynamic
Turbulence: Results from a One-Dimensional Model}  
\author{Abhik Basu$^1$, Jayanta K Bhattacharjee$^2$, 
and Sriram Ramaswamy$^1$}
\address{$^1$Department of Physics, Indian Institute of Science, 
Bangalore 560012 INDIA, and $^2$Indian Association for
the Cultivation of Science, Calcutta 700032,
INDIA} 
\maketitle
\begin{abstract}
We show that a recently proposed [J. Fleischer and P.H. Diamond, 
{\em Phys. Rev. E}{\bf 58}, R2709 (1998)] one-dimensional 
Burgers-like model for magnetohydrodynamics (MHD) is in effect
identical to existing models for drifting lines  
and sedimenting lattices.  
We use the model
to demonstrate, contrary to claims in the literature, that the 
energy spectrum of MHD turbulence 
should be independent of mean magnetic field  
and that cross-correlations between the noise sources
for the velocity and magnetic fields cannot
change the structure of the equations under renormalisation. 
We comment on the scaling and the multiscaling properties
of the stochastically forced version of the model. 
\end{abstract}


\section{Introduction and Results}
The search for simple model equations embodying
some of the features of fluid turbulence prompted
Burgers to propose his famous nonlinear diffusion
equation \cite{burgers}
\begin{equation}
{\partial u\over\partial t} + u
{\partial\over\partial x}u =
\nu_o {\partial ^2 u\over\partial x^2} + f(x,t) 
\label{burg}
\end{equation}
in one space dimension, where $\nu_0$ is a
``viscosity'' and $f$ a forcing function. The
unique properties of (\ref{burg}), particularly
the Cole-Hopf transformation \cite{burgers} that
linearises it, together with its obvious formal
similarity to the Navier-Stokes equation, have
led to many studies \cite{burgulence,hayot} shedding
some light on real fluid turbulence. Turbulence
in {\em plasmas} is believed to be described by
the equations of magnetohydrodynamics \cite{mont} in three
dimensions (3$d$MHD) for the coupled
evolution of the velocity field ${\bf u}$ and the
magnetic field ${\bf B}$. In 3$d$MHD, the  
Navier-Stokes equation for an incompressible fluid is modified by the
inclusion of electromagnetic stresses: 
\begin{equation}
{\partial {\bf u}\over \partial t}+ {\bf
u}.\nabla {\bf u}=-{\nabla p\over \rho} +{(\nabla
\times {\bf B})\times {\bf B}\over\ 4\pi \rho} +\nu \nabla^2 {\bf u}
+{\bf f}_u 
\label{mhdu}
\end{equation}
with $\nabla \bf .u=0$; and Amp\`ere's law for a conducting 
fluid becomes
\begin{equation}
{\partial {\bf B}\over\partial t}+{\bf u.\nabla
B}={\bf B.\nabla u} +\mu \nabla^2 {\bf B} +
{\bf f}_b.
\label{mhdb}
\end{equation}
In (\ref{mhdu}) and (\ref{mhdb}), $\rho$ is the fluid
density, $p$ is the pressure, $\mu$ is the ``magnetic
viscosity'', arising from the nonzero resistivity
of the plasma, $\nu$ is the
kinematic viscosity and $\bf f_u$ and $\bf f_b$ are forcing functions. 
Since the MHD equations are distinctly more complicated
than the Navier-Stokes equation, a  toy model, whose relation to 
the MHD equations is the same as that of the Burgers equation 
to the Navier-Stokes equation, should be very welcome. 

The first attempt in this direction \cite{thomas} lacked cross-helicity  
conservation. The $1d$ model with all the {\em scalar} 
conservation laws of $3d$MHD is that   
due to Fleischer and Diamond \cite{flei}. Their  
equations, however, have been in the 
literature for some time now in a different context. 
For the case where the mean magnetic 
field $b_0 = 0$, the equations are those of the 
Erta\c{s}-Kardar \cite{er} model of drifting lines. For the general 
case $b_0 \neq 0$, they are the coupled equations, for the displacements  
along and normal to the drift direction, of the one-dimensional reduced 
model of Lahiri and Ramaswamy \cite{rlsr} for a crystalline lattice 
moving through a dissipative medium. 
This surprising relation arises from an exact correspondence between
the symmetries of these equations and those of the $1d$MHD
equations \cite{flei} with and without $b_0 = 0$.
The relation between the velocity and magnetic fields in \cite{flei} 
and the displacement fields in \cite{er,rlsr} is precisely 
the same as that between the height in the KPZ equation \cite{kpz} 
and the velocity in (\ref{burg}).  

Reduced models are useful if they allow one 
to answer a question of principle. We use the $1d$MHD model to show that
the form of the energy spectrum in MHD with a mean magnetic field
should be the same as that without one, contrary to a claim by Kraichnan
\cite{krai}. We also explore the scaling and multiscaling
properties of the model equations in the presence of stochastic forcing. 
For a temporally white noise source with variance 
$\propto k^{-y + 3}$ at small wavenumbers 
$k$, we find ``sweeping'' divergences for $y \geq 3$ and, 
within perturbative and renormalisation-group analyses, we show that 
the effective 
kinematic and magnetic viscosities are equal at long wavelengths. 
Specifically for $y = 4$, 
as is appropriate \cite{burgulence,hayot} for a study of $1d$ turbulence, we find within a 
scaling treatment that   
the energy spectrum $E(k) \sim k^{-5/3}$ and the dynamic 
exponent $z = 2/3$. 
We show also 
that the multiscaling properties of the model should be precisely 
the same as that of a Burgers equation with the same driving \cite{hayot}.
Lastly, we correct a claim \cite{cam} in the literature 
regarding the role of cross-correlations between the noise 
sources in the velocity and magnetic-field equations.  

\section{Symmetries, equations of motion, and noise statistics}
Let us remind the reader of the ``derivation''
of the $1d$MHD equations \cite{flei} emphasising some points
which we feel are important and are missing in ref. \cite{flei}.
The structure of the $1d$ model is determined by
the invariances of the MHD equations.  (i) The
velocity and magnetic fields are polar and
axial vectors respectively. For a model in one space
dimension with one-component fields $u(x)$ for the 
velocity and $b(x)$ for the magnetic field 
this implies evolution equations which are invariant
under $x \rightarrow -x$ together with $u
\rightarrow -u$, with $b \rightarrow b$. (ii) The MHD equations
are Galileian-invariant, so our $1d$ model must
be unchanged under $x = x' + u_0 t', \, t = t',\,
u(x,t) = u'(x',t') + u_0,\,b(x,t)=b'(x',t')$ for any constant $u_0$.
(iii) The $1d$ equations, like the $3d$
originals, must have the ``conservation law''
form $\partial_t(\,\,) = \nabla . (\,\,)$.  
(iv) The equations, in the absence of the diffusion
terms, must conserve the ``energy'' 
\begin{equation}
\label{energy}
{\cal E} \equiv \int (u^2+\beta b^2) dx
\end{equation}
for some constant $\beta$.
To leading orders in gradients and bilinear
order in fields, the most general $1d$ MHD model
consistent with these requirements \cite{flei} is \footnote{The choice of
bookkeeping coefficients in the equations as presented in \cite{flei}
seems unnatural to us. Galileian invariance requires equality
of the coefficients of $u\partial_x u$ and $\partial_x (ub)$,
while that of $b\partial_x b$ can be left entirely arbitrary.}
\subequations
\begin{eqnarray}
{\partial u\over\partial t} + 
\lambda_2 b_0 {\partial b\over
\partial x} +  \lambda_1
u{\partial u\over\partial x} +
\lambda_2 b{\partial b\over
\partial x} &=& \nu_o {\partial ^2 u\over \partial
x^2} + f_u,  
\label{mhd1du} \\
{\partial b\over\partial t} +  \lambda_1 b_0 
{\partial\over\partial x} u   + \lambda_1
{\partial\over\partial x} (ub) &=& \mu_o {\partial
^2 b\over
\partial x^2} + f_b.
\label{mhd1db}
\end{eqnarray}
\label{mhd1d}
\endsubequations
Integrating (\ref{mhd1d}) once with respect 
to $x$, one finds that $\int_x u$ and $\int_x b$ obey  
the equations of Erta\c{s} and Kardar \cite{er} for drifting 
lines when $b_o = 0$, and those of Lahiri and Ramaswamy \cite{rlsr} 
for sedimenting lattices  when $b_o \neq 0$. This transformation 
of variables is the same as that which takes one from the Burgers 
equation to the KPZ \cite{kpz} equation.  

In (\ref{mhd1d}), the constant $b_o$ is the ``mean magnetic field'', 
$\nu$ and $\mu$ are the kinematic and magnetic viscosities, 
and $f_u$ and $f_b$ are forcing
functions.  $\lambda_1$ can always be rescaled to
unity but is retained for book-keeping purposes. $\lambda_2$ is 
arbitrary, and all statements made in this paper hold independent of 
its value \cite{flei}. 
It is straightforward to show that the equations, in the absence of
the diffusion terms, conserve the energy-like quantity ${\cal E}$ 
defined in  eq.(\ref{energy}), for $\beta=\lambda_2/\lambda_1$, as well as  
$\int_x ub$, the analogue in $d = 1$ of the cross helicity
($\int_x {\bf u}.{\bf B}$) \cite{mont}, and that the model of 
\cite{thomas} does not conserve the latter. A third conserved quantity, 
the magnetic helicity $\int {\bf A}.{\bf B}$, where $\bf A$ is the 
vector potential, has no analogue in $d = 1$. 
It is important to note that for $b_o$=0  
the equations have a higher symmetry than for $b_0 \neq 0$. For 
both $b_0 = 0$ and $b_0 \neq 0$, they are of course invariant under 
$x\rightarrow -x, u\rightarrow -u$, $b\rightarrow b$, as 
imposed by the transformation properties of the fields. For 
$b_0 = 0$, they are {\em in addition} invariant under 
$x\rightarrow -x, u\rightarrow -u$, $b\rightarrow -b$. 
For $\lambda_1\lambda_2 > 0$, in the linear approximation 
at small wavenumber $k$, as discussed in \cite{rlsr}, the 
equations (\ref{mhd1d}) have a wavelike response with 
frequency $\omega=\pm \sqrt{\lambda_1 \lambda_2 b_o ^2}$.  
These are the $1d$ analogue of Alfv\'{e}n waves \cite{mont,flei}. 
The linearly unstable case
$\lambda_1\lambda_2 <0$ \cite{rlsr} will not be considered here  
since the corresponding possibility is unphysical in the 
context of the $3d$MHD equations (\ref{mhdu}) and (\ref{mhdb}).  

Since we wish to study turbulence in (\ref{mhd1d}), 
we shall take the forcing terms $f_u, \, f_b$ to be 
zero-mean Gaussian noise sources with covariance specified 
below. The properties of $u$ and $b$ under $x\rightarrow -x$ 
and the reality of $f_u(x,t)$ and $f_b(x,t)$ imply that 
$<f_u(k,0)f_u(-k,t)>$ and $<f_b(k,0)f_b(-k,t)>$ are real and even 
in $k$ while, crucially, $<f_u(k,0)f_b(-k,t)>$ (if nonzero) is odd in $k$ 
and purely imaginary. In general, therefore, 
\subequations
\begin{eqnarray}
<f_u(k,0)f_u(k',t)> &=&  A_u(|k|) \delta (k+k^{\prime}) \delta(t)
\label{noiseu}
\\
<f_b(k,0)f_b(k',t)> &=&  A_b(|k|) \delta (k+k^{\prime}) \delta(t)
\label{noiseb}
\\
<f_u(k,0)f_b(-k,t)> &=& i k C(|k|) \delta (k+k^{\prime}) \delta(t), 
\label{noiseub}
\end{eqnarray}
\label{noisebub}
\endsubequations 
where $A_u, \, A_b$ and $C$ are {\em real} functions. 
By strict analogy with the random stirring approach in $d=3$ 
\cite{dedom,orz} and $1d$ Burgers turbulence \cite{hayot} we choose
\begin{equation}
\label{epsu}
A_u(k) = {\epsilon_u \over |k|} 
\end{equation}
\noindent so that the noise strength $\epsilon_u$ has the units 
of the dissipation parameter, i.e. (length)$^2$/(time)$^3$. 
$\langle f_u f_b \rangle$ and $\langle f_b f_b \rangle$ can be taken to 
be zero or at any rate no more singular than $\langle f_u f_u \rangle$ 
without altering any of our main conclusions. A self-consistent treatment
starting with only a nonzero $\langle f_uf_u \rangle$ will generate $\langle f_uf_b \rangle$
and $\langle f_b f_b \rangle$ as well. In that sense, contrary to the statements
in \cite{flei}, the case where only one of the equations has
a bare noise is not physically distinct from that where both noises
are non-zero.
For convenience in carrying out bare perturbation theory, 
however, we take all components of the noise covariance 
to be as singular as $\langle f_u f_u \rangle$:  
\begin{eqnarray}
A_b(|k|) = {\epsilon_b \over |k|} 
\label{epsb}
\\
k C(|k|) = {\epsilon_{ub} \over k} 
\label{epsub}
\label{epsbub}
\end{eqnarray}
Note that it is $1/k$ and not $1/|k|$ that appears on the 
right-hand side of (\ref{epsub}).
We now use the reduced equations (\ref{mhd1d}) to obtain 
the results advertised in the Introduction.

\section{Energy spectrum and the role of a mean magnetic field}

Does the presence of a mean magnetic field $b_0$ affect 
the energy spectrum $E(k)$ at wavenumber $k$ in MHD turbulence? 
It has been argued \cite{krai} that if $b_0 \neq 0$ 
then (i) the dissipation $\epsilon$ must be proportional 
to the Alfv\'en wave propagation time $1/c k$, 
where $c$ is the wavespeed; (ii) $\epsilon$  must depend, 
apart from this, only on $k$ and $E(k)$; and (iii) as 
a result of energy conservation and a local-cascade picture, 
$\epsilon$ must be independent of $k$. Taken together, these 
imply \cite{krai} that $E(k) \sim k^{-3/2}$, while for $b_0 =0$ 
the usual Kolmogorov arguments yield $k^{-5/3}$. 

The questionable 
assumption in the foregoing analysis is that the wave-propagation 
time sets the scale for the dynamics. 
In what follows, we test that assumption in the $1d$MHD (\ref{mhd1d}) 
model since the above arguments, 
if correct, should apply there as well.  
We show below explicitly that the presence of a nonzero wavespeed 
leaves the equal-time correlations unchanged at small $k$. 
We begin by rewriting (\ref{mhd1d}) in terms of 
the Els\"asser \cite{els} variables 
\begin{equation}
\label{z+-}
z^{\pm}={u \pm \sqrt \lambda_2 b\over\sqrt{1+\lambda_2}}. 
\end{equation}
As has already been noted in \cite{er,rlsr}, the equations for $z^+$ 
and $z^-$ decouple completely \cite{flei} at 
small wavenumbers \footnote{The decoupling is demonstrated in \cite{flei}
for zero mean magnetic field, with equal magnetic and kinematic 
viscosities. Our treatment is somewhat more general.}: 
\begin{mathletters}
\begin{eqnarray}
{\partial z^+\over\partial t} - b_o\partial_x z^+ + 
{1\over\ 2}{\sqrt{1+\lambda_2}\over\ 2}
\partial_x {z^+}^2=D\partial_{xx}z^+ + \ldots + f^+ ; 
\label{eqz+}
\\
{\partial z^-\over\partial t} + b_o\partial_x z^- +  
{1\over\ 2}{\sqrt{1+\lambda_2}\over\ 2}
\partial_x {z^-}^2=D\partial_{xx}z^- + \ldots + f^-, 
\label{eqz-} 
\end{eqnarray}
\label{eqz+-}
\end{mathletters}
\noindent with
\begin{equation}
\label{f+-}
f^{\pm}={f_u \pm \sqrt \lambda_2 f_b\over\sqrt{1+\lambda_2}},  
\end{equation}
where we have set $\lambda_1$ to unity, $D=(\mu+\nu)/2$,
and the ellipsis 
refers to reactive (i.e., nondissipative) 
terms proportional to $\nu - \mu$, subdominant in wavenumber 
relative to the leading Alfv\'en wave terms $\propto b_0$. 
It therefore suffices to study the case $\nu = \mu = D$. 
In (\ref{eqz+}) and (\ref{eqz-}), the wave terms 
$\propto b_0$ can be absorbed by opposite Galileian boosts, 
$x \rightarrow x - b_0 t$, and $x \rightarrow x + b_0 t$ 
respectively. Each equation can therefore be studied independently,  
at small wavenumber, as far as the autocorrelations of $z^+$ or 
$z^-$ are concerned. 
However, the correlations of the physical fields $u$ and $b$ 
involve correlations $\langle z^+ z^- \rangle$ (which 
are nonzero because $\langle f^+ f^- \rangle \neq 0$). 
Some care must be taken while evaluating these, as we will 
show in detail below.
Since 
\begin{equation}
u = {\sqrt{1 + \lambda_2} \over 2}(z^+ + z^-) 
\label{uz}
\end{equation} 
and 
\begin{equation}
b = {\sqrt{1 + \lambda_2} \over 2 \sqrt{\lambda_2}}(z^+ - z^-) 
\label{bz}
\end{equation}
we see that 
\begin{eqnarray}
\langle u(k,t) u(-k,0) \rangle = {{1 + \lambda_2} \over 4}
[\langle z^+(k,t)z^+(-k,t)\rangle +\langle z^-(k,t)z^-(-k,t)\rangle
+2Re(\langle z^+(k,t)z^-(-k,0)\rangle )] \\
\langle b(k,t) b(-k,0) \rangle = {{1 + \lambda_2} \over 4\lambda_2}
[\langle z^+(k,t)z^+(-k,t)\rangle +\langle z^-(k,t)z^-(-k,t)\rangle
-2\mbox{Re}(\langle z^+(k,t)z^-(-k,0)\rangle )] 
\end{eqnarray}
and
\begin{equation}
\langle u(k,t)b(-k,0)\rangle={{1 + \lambda_2} \over 4\sqrt \lambda_2}
[\langle z^+(k,t)z^+(-k,t)\rangle-\langle z^-(k,t)z^-(-k,t)\rangle
-2\mbox{Im}(\langle z^+(k,t)z^-(-k,0)\rangle )]
\end{equation}

Let us define shifted coordinates 
\begin{equation}
y_{\pm} = x \mp b_o t
\label{y+-}
\end{equation}
and ``comoving'' fields
\begin{equation}
\zeta^{\pm}(y_{\pm},t)\equiv\zeta^{\pm}(x\mp b_ot,t)=z^{\pm}(x,t). 
\end{equation}
Then, Fourier transforming, we get
\begin{equation}
z^{\pm}(k,t)=e^{\mp ikb_ot}\zeta^{\pm}. 
\end{equation}
The autocorrelations of $z^+$ and $\zeta^+$, as 
well as those of $z^-$ and $\zeta^-$, are   
related by travelling waves:  
\begin{equation}
\label{zcorrpmpm}
\langle z^{\pm}(k,t)z^{\pm}(k',t')\rangle =
e^{\mp i kb_o(t-t')}\langle \zeta^{\pm}(k,t)\zeta^{\pm}(k',t')\rangle,  
\end{equation}
while that between $z^+$ and $z^-$ has a phase factor 
which depends on the {\em sum} of the time arguments: 
\begin{equation}
\label{zcorr+-}
\langle z^+(k,t)z^-(k',t')\rangle =
e^{-i kb_o(t-t')}\langle \zeta^+(k,t)\zeta^-(k',t')\rangle.   
\end{equation}
Defining 
\begin{equation}
\label{lamtild}
\tilde{\lambda} \equiv {\sqrt{1 + \lambda_2} \over 2}, 
\end{equation}
we see that the fields $\zeta^{\pm}$ obey the Burgers equation
\begin{equation}
\partial_t \zeta_{\pm}+{\tilde{\lambda}\over\ 2}\partial_y \zeta_{\pm}^2 
=D\partial_{yy} \zeta_{\pm}+\phi_{\pm}(y,t)
\label{zetabur}
\end{equation}
with noise correlations related to those of $f^+$ and $f^-$ 
in eq. (\ref{f+-}) by 
\begin{equation}
\label{phi+phi+}
\langle \phi^+(k,t) \phi^+(k',t') \rangle = 
\langle f^+(k,t) f^+(k',t') \rangle,  
\end{equation}
\begin{equation}
\label{phi-phi-}
\langle \phi^-(k,t) \phi^-(k',t') \rangle = 
\langle f^-(k,t) f^-(k',t') \rangle,  
\end{equation}
and 
\begin{equation}
\label{phi+phi-}
\langle \phi^+(k,t) \phi^-(k',t') \rangle = 
e^{i k b_0 (t+t')} \langle f^+(k,t) f^-(k',t') \rangle.  
\end{equation}
Despite the appearance of nonstationary phase factors in 
the correlations $\langle \phi^+ \phi^- \rangle$ and 
$\langle \zeta^+ \zeta^- \rangle$, as a result of 
the time-dependent coordinate transformation,  
all physical correlation functions
will of course be time-translation-invariant. 
The autocorrelations $\langle \zeta^+ \zeta^+ \rangle$ and
$\langle \zeta^- \zeta^- \rangle$ are particularly 
simple and entirely independent of the wavespeed $b_0$.  
Thus the energy 
spectrum which, for $\lambda_1 = 1$, is 
\begin{eqnarray}
\label{espec}
E(k) &=& {1 \over 2} \int {\mbox{dk} \over 2 \pi} 
\left [\langle |u(k)|^2 \rangle +  
\lambda_2 \langle |b(k)|^2 \rangle \right ] \\   
&=& {{1 + \lambda_2} \over 2} 
\int {\mbox{dk} \over 2 \pi} 
\left [\langle |z^+(k)|^2 \rangle +  
\langle |z^-(k)|^2 \rangle \right ] \\   
&=& {{1 + \lambda_2} \over 2} 
\int {\mbox{dk} \over 2 \pi} 
\left [\langle |\zeta^+(k)|^2 \rangle +  
\langle |\zeta^-(k)|^2 \rangle \right ]    
\end{eqnarray}
is therefore {\em identical} to that of the randomly 
forced Burgers equation (\ref{zetabur}), which does not 
contain the Alfv\'{e}n waves at all. This refutes the claim of 
\cite{krai}, since the arguments in that work would, if correct, 
have applied to the present model as well. This establishes 
the assertion in the Introduction that a nonzero mean 
magnetic field is irrelevant to a determination of 
the energy spectrum of MHD turbulence. It is worth noting 
that our result holds regardless of the nature of the 
forcing (deterministic, stochastic, with or without 
long-range correlations). 

A demonstration that  
{\em all} equal-time correlations in (\ref{mhd1d}) are 
independent of $b_0$, requires  
showing that $\langle z^+(k) z^-(-k) \rangle\/$ does 
not involve the wavespeed. We establish this below  
for sufficiently small wavenumber. We must assume in our 
derivation that the long-wavelength behaviour of (\ref{zetabur}) 
is determined by a fixed point at which Galileian-invariance 
prevents the nonlinear coupling in (\ref{zetabur}) from renormalising, 
and at which higher-order nonlinearities are not relevant. 
In that 
case, the {\em renormalised} $\langle z^+(k) z^-(-k) \rangle\/$ correlation 
function can be written in terms of the  
{\em renormalised} propagator $G(k,t)$ of the randomly-forced Burgers 
equation (\ref{zetabur}) and the (renormalised, in principle) correlator 
$N_{+-}(k,t)$ of $f^+(k,t)$ with $f^-(-k,0)$ with appropriate 
phase factors arising from the change of variables (\ref{y+-}):  
\begin{eqnarray}
\label{S+-kt}
\langle z^+(k,t)z^-(k',t')\rangle &=&
e^{-i k b_o t + i k' b_o t'} \delta (k+k') \int_{-\infty}^t dt_1 
\int_{-\infty}^{t'} dt_2 
 G(k,t-t_1) N _{+-}(k,t_1 - t_2)
e^{ib_ok(t_1 +  t_2)}G (-k,t'-t_2) \\ 
&\equiv& S_{+-}(k,t-t')\delta(k+k')
\end{eqnarray}
Redefining $t - t_1 \rightarrow t_1$, $t' - t_2 \rightarrow t_2$, we 
obtain the equal-time correlations  
\begin{equation}
S_{+-}(k) = \int_0 ^{\infty} dt_1 \int_0 ^{\infty} dt_2 
G(k,t_1)G(-k,t_2)e^{i k b_o (t_1 + t_2)} N_{+-}(k, t_2 - t_1)
\label{S+-k}
\end{equation}
Scaling tells us that 
\begin{equation}
G(k,t)=\Gamma (k^zt)
\label{Gamma}
\end{equation}
and 
\begin{equation}
N_{+-}(k,t)=k^z R_{+-}(k)\Gamma_N(k^zt)
\label{GammaN}
\end{equation}
where $z$ is the dynamic exponent for equations (\ref{zetabur}), i.e., 
without the wave terms, $\Gamma$ and $\Gamma_N$ are scaling functions, 
and $R_{+-}(k)$ is the zero-frequency renormalised covariance 
of $f^+$ and $f^-$. Defining $\tau_1 = k^z t_1$ and  
$\tau_2 = k^z t_2$, we see that  
\begin{equation}
S_{+-}(k) = R _{+-}(k)\int_0 ^{\infty} d \tau_1 \int_0^{\infty} d \tau_2
 \Gamma(\tau_1) \Gamma(\tau_2) \Gamma_N(\tau_1 - \tau_2)  
e^{i(\tau_1 + \tau_2) b_o k^{1-z}}.    
\label{S+-k2}
\end{equation}
Since the dynamic exponent governing the decay of correlations in 
the Burgers equation with noise variance $\sim 1/k$ is known 
\cite{hayot,burgulence} to be 2/3, the phase factor inside the integral
(\ref{S+-k2}) for small $k$ can be set to unity, i.e., the wavespeed 
term drops out. This completes the demonstration 
that equal-time correlations in MHD turbulence are independent 
of the mean magnetic field\footnote{Even if we had a model with $z > 1$, say, 
for a noise with vanishing variance at zero wavenumber, the contribution 
from (\ref{S+-k2}) would be of the form (noise-strength/$b_ok$), while 
those from the $z^+ z^+$ and  $z^- z^-$ would be of order 
(noise-strength/$k^z$). The former, while in principle dependent 
on the wavespeed, would be subdominant (since we are now considering 
$z > 1$) to the latter.}.   

\section{Scaling, renormalisation group and self-consistent 
analyses for zero mean magnetic field}

Some general  observations will allow us to estimate the scaling exponents
for this $1d$MHD model. First the Galileian invariance mentioned before 
implies that $\lambda_1$ will not renormalise. Secondly, the fact that $b_o$
can be eliminated completely from equation (\ref{z+-}) to give 
equations (\ref{zetabur}) by the transformations $x\rightarrow x\pm b_o t$
means that $b_o$ cannot renormalise. Thirdly, the invariance of equations
(\ref{zetabur}) under the Galileian transformation $\zeta^{\pm}(x,t)=
\zeta^{\pm}(x+\lambda_2 u_ot,t)\pm u_o,$
 means that the 
coupling strength $\lambda_2$ in equations 
(\ref{zetabur}) cannot renormalise \footnote{This is another point 
of similarity with $3d$MHD: In terms of the Els\"{a}sser \cite{els} 
variables ${\bf z}^{\pm}$, the nonlinearity in the $3d$MHD equation  
for ${\bf z}^+$ is  ${\bf z}^- . {\bf \nabla z}^+$, and that 
in the equation for ${\bf z}^-$ is  ${\bf z}^+ . {\bf \nabla z}^-$. 
The equations in this form have the Galileian invariance ${\bf z}^{\pm} 
\rightarrow  {\bf z}^{\pm} + {\bf u}_o, \, {\bf r} \rightarrow 
{\bf r} - {\bf u_o}t$ for any constant vector ${\bf u}_o$, 
where ${\bf r}$ is the position coordinate. This guarantees that all 
the vertices in $3d$MHD are unaffected by fluctuation-corrections at 
zero wavenumber.}.

Thus, if we were to carry out a renormalisation-group
transformation by integrating out a shell of modes 
$\Lambda e^{-\ell}<q<\Lambda$, and rescaling $x \rightarrow e^{\ell} x, \,  
u \rightarrow e^{\ell \chi_u} u, \, b \rightarrow e^{\ell \chi_b} b, \, 
t \rightarrow e^{\ell z} t)$, 
the couplings $\lambda_1$ and $\lambda_2$ would be affected
only by the rescaling:
\begin{equation}
\lambda_1\rightarrow e^{\ell (\chi_u+z-1)}\lambda_1, 
\lambda_2\rightarrow e^{\ell (2\chi_b-\chi_u+z-1)}\lambda_2.
\end{equation}
We can thus rescale to keep $\lambda_1$ and $\lambda_2$
fixed, giving 
\begin{equation}
\chi_u=\chi_b=1-z. 
\end{equation}
The noises sources $f_u$ and $f_b$ have variances (see equation(\ref{epsbub}))
which diverge at small wavenumber $k$. Since the nonlinear couplings are first
order in $k$, any renormalisation of $\epsilon_u, \epsilon_b$ and $\epsilon_{ub}$
will be of the form $k^2W(k,\omega)$ which vanishes for $k\rightarrow 0, \omega \neq 0$.
The noise is thus unrenormalised at small $k$ for any nonzero frequency.
If we extend this to say that the noise strength receives no fluctuation
corrections at all, then, under a renormalisation-group transformation, the
noise strength too is affected only by rescaling. Insisting that the rescaling
leave the noise strength unchanged yields $z=2\chi_u=2\chi_b$. Thus, $z=2/3$
and $\chi_u=\chi_b=1/3$, so that $E(k)\sim k^{-5/3}$. While this treatment,
which is the same as that applied to the Burgers equation with noise
variance $1/k$, does not yield multiscaling, it appears \cite{hayot} to
be satisfactory for two point correlations. Since our equations
for $\zeta^{\pm}$ (\ref{zetabur}) are identical to \cite{hayot}, 
it is reasonable to expect that the $1d$MHD equations (\ref{mhd1d}) 
with singular noise will show multiscaling similar to 
that in \cite{hayot}. In particular, in this $1d$ treatment, 
$z^+$ and $z^-$ obey the same equation
implying that their multiscaling properties are the same. This is consistent
with the behaviour of the Els\"{a}sser fields in a shell model 
of $3d$MHD \cite{abhik}.

In section (III) our decoupling of $z^{+}$ and $z^{-}$ required $\mu=\nu$.
For $b_o \neq 0$, we showed that terms involving $\mu-\nu$ were subdominant
to the $b_o$ terms. Since the same arguments cannot be made for
$b_o$ =0, we have carried out a dynamic renormalisation group treatment in
the absence of a mean magnetic field, following the momentum shell
approach of \cite{ma,fns}. In our treatment we allow for independent
coupling terms $\lambda_1u\partial_x u$, and $\lambda_2 b\partial_x b$ in 
the $u$ equation and $\lambda_3 \partial_x(ub)$ in the $b$ equation.
We assume, as in the scaling argument above, that there is no 
diagramatic correction to the noise strength, and we choose the 
rescaling so that the noise strength
remains fixed. For simplicity we ignore the cross-correlation 
$\langle f_u f_b\rangle$. The resulting recursion relations have a stable 
fixed point at which $\mu=\nu$. 
This allows us to work with $\mu = \nu$, and hence to use the 
decoupled description even when $b_o=0$.

We have also carried out a one-loop self-consistent treatment of
a slight generalisation of equations (\ref{mhd1d}) and (\ref{noisebub}) 
with $A_u(|k|) \sim D_1k^{-y+3}, A_b(|k|) \sim D_2k^{-s+3}$ and $kC(|k|)=0$. 
Ignoring noise and vertex renormalisations, we find that the fluctuation
corrections $\Sigma_u (k)$ and $\Sigma_b (k)$ at zero frequency
for $\nu$ and $\mu$ respectively obey:
\begin{eqnarray}
\Sigma^u(k) &=& \int {dp\over\ 2\pi} 
\left\{{D_1 p^{-y+3} \over {p^2\Sigma^u(p) 
\left[p^2\Sigma^u(p) + (k-p)^2 \Sigma^u (k-p)\right]}} \right. + \\ 
&&\left.{D_2 p^{-s+3} \over {p^2\Sigma^b(p)  
\left[p^2\Sigma^b(p) + (k-p)^2 \Sigma^b (k-p)\right]}}\right\} 
\label{sigu}
\end{eqnarray}
and 
\begin{eqnarray}
\Sigma^b(k) &=& \int {dp\over\ 2\pi} 
\left\{{D_1 p^{-y+3} \over {p^2\Sigma^u(p) 
\left[p^2\Sigma^b(p) + (k-p)^2 \Sigma^b (k-p)\right]}} \right. + \\ 
&&\left.{D_2 p^{-s+3} \over {p^2\Sigma^b(p)  
\left[p^2\Sigma^u(p) + (k-p)^2 \Sigma^u (k-p)\right]}}\right\}.  
\label{sigb}
\end{eqnarray}
If we seek a solution of the form $\Sigma_u =\Gamma_u k^{x_1}$ and $\Sigma_b
=\Gamma_b k^{x_2}$ we find consistency requires $x_1=x_2=y/3=s/3$ and $\Gamma_u 
=\Gamma_b$, i.e., that the scale-dependent kinematic and magnetic 
viscosities are identical at small $k$. The singularities of the one-loop integral depend strongly
upon $y$. For $y\geq 3$, these integrals diverge even when the external 
wavenumber $k\neq 0$ (the ``sweeping divergence'' \cite{jkb,mou}).
For $y$ near 0, on the other hands, one finds  
$\Sigma_u (k)\sim\Sigma_b (k)\sim k^{-y/3}$. 
These singularity structures are very similar to those seen in 
a self-consistent treatment of the randomly stirred Navier-Stokes equations. 

A technical clarification may be of some
interest here. In carrying out the perturbation theory with $b_o=0$, 
care must be taken to ensure that the condition $\langle b \rangle =0$ 
is maintained
order by order. If $ikC(|k|)\neq 0$, i.e., there are
cross correlations in the noise sources in equations (\ref{mhd1d}),  
and $\mu\neq\nu$, then  
the perturbation theory generates an apparent non-zero $<b>$
and an apparent Alfv\'{e}n wave speed. Including a 
counterterm to cancel the spurious $<b>$ automatically
ensures the absence of Alfv\'{e}n waves at $b_o=0$. At the RG fixed point
discussed above, where $\mu=\nu$, this issue clearly does not arise.

Our final point concerns the assertion
\cite{cam} that the introduction
of a nonzero cross-correlation $\langle f_u f_b \rangle $
in $3d$MHD generates, under renormalisation, dissipative terms of the
form $\nabla^2 {\bf B}$ in the equation for ${\bf u}$ and
$\nabla^2 {\bf u}$ in the equation for $\bf B$. If this were
true, the renormalised equations of motion would lack the symmetry
properties enjoined on them by the fact that $\bf B$ and $\bf v$
are a pseudovector and a vector respectively. Unsurprisingly,
a straightforward perturbative analysis 
of fluctuation corrections arising from the nonlinearities
in (\ref{mhd1d}) rules out any such anomaly. As long as the statistics of
the noise sources are as in (\ref{noiseu}) and (\ref{noiseub}),
so that the transformation properties of $v$ and $b$ are respected,
no such terms are generated. It is clear that the error
in \cite{cam} arose because their cross-correlation
$\langle f_u(k) f_b(-k) \rangle$ was {\em real} and {\em even}
in wavevector, which is simply not consistent with the nature of the fields
in the problem.  Such a cross-correlation will in fact generate not only
the terms mentioned above but terms like $\partial_x (ub)$ in the $u$
equation and $\partial_x(u^2)$ and $\partial_x(b^2)$ in the $b$
equation as well. The resulting system will have only the invariance
$x\rightarrow -x, u\rightarrow -u$, $b\rightarrow -b$, which has
nothing to do with the intrinsic transformation properties of the
physical fields.

\section{summary}
We have shown that a recently proposed \cite{flei} Burgers-like 
one-dimensional model
for magnetohydrodynamics ($1d$MHD)  is completely equivalent, through a 
simple transformation,  
to existing equations in the literature \cite{er,rlsr}. We have obtained 
several new results from these $1d$MHD equations.
The most important of these, which should apply to $3d$MHD as well,
is that the energy spectrum
is unaffected by the presence of a mean magnetic field.
Apart from this, we have presented scaling, renormalisation-group
and self-consistent analyses of the large-distance, long-time
behaviour of correlation functions in the model, and provided
arguments for the existence of multiscaling therein when
the forcing functions have singular small-wavenumber correlations.
Lastly, we have corrected an erroneous claim \cite{cam} regarding 
the effect of cross-correlations between the forcing functions
in the velocity and the magnetic field equations.

\section{Acknowledgement}
We thank Rahul Pandit for useful discussions and a critical 
reading of the manuscript. AB was supported by the Council 
for Scientific and Industrial Research, India.


\begin{references}
\bibitem{burgers} J.M. Burgers, {\it The Nonlinear Diffusion Equation}, 
(Reidel, Boston, 1977)
\bibitem{burgulence} A. Cheklov and V. Yakhot, Phys. Rev. E {\bf 52}, 
5681 (1995). 
\bibitem{hayot} F. Hayot and C. Jayaprakash, {\em Phys. Rev. E} {\bf 54},
4681 (1996). 
\bibitem{mont} For a review, see D. Montgomery in {\em Lecture 
Notes on Turbulence}, eds.  J. R. Herring and J. C. McWilliam 
(World Scientific, Singapore, 1989).
\bibitem{thomas} J.H. Thomas, {\em Phys. Fluids} {\bf 11}, 1245 (1968).  
\bibitem{flei} J. Fleischer and P.H. Diamond, {\em Phys. Rev. E} 
{\bf 58}, R2709 (1998); this paper came to our notice after our 
work was complete, and while our manuscript was being typed.
\bibitem{er} D. Ertas and M. Kardar, {\em Phys. Rev. E}{\bf 48}, 
1228 (1993).
\bibitem{rlsr} R. Lahiri and S. Ramaswamy, {\em Phys. Rev. Lett.} {\bf 79},
1150 (1997). 
\bibitem{kpz} M. Kardar, G. Parisi, and Y.-C. Zhang, Phys. Rev. Lett. 
{\bf 56}, 889 (1986). 
\bibitem{krai} R. H. Kraichnan, {\em Phys. Fluids}, {\bf 8}, 1385 (1965).
\bibitem{cam} S. J. Camargo and H. Tusso, {\em Phys. Fluids} {\bf B} {\bf 4},
1199 (1992). 
\bibitem{dedom} De Dominicis and P. C. Martin, {\em Phys. Rev. A} {\bf 19},
419 (1979).
\bibitem{orz} V. Yakhot and S. A. Orzag, {\em Phys. Rev. Lett.} {\bf 57},
1722 (1986).
\bibitem{els} W.M. Els\"{a}sser, {\em Phys. Rev.} {\bf 79}, 183 (1950).  
\bibitem{abhik} A. Basu, A. Sain, S. K. Dhar and R. Pandit,
{\em Phys. Rev. Lett.} {\bf 81}, 2687 (1998).
\bibitem{ma}S. K. Ma and G. F. Mazenko, {\em Phys. Rev. B} {\bf 11}, 4077 
(1975).
\bibitem{fns} D. Forster, D. R. Nelson, and M. J. Stephen, 
{\em Phys. Rev. A} {\bf 16}, 732 (1977).
\bibitem{jkb} J. K. Bhattacharjee, {\em J. Phys. A. Math. Gen.} {\bf 21},
L551 (1988).
\bibitem{mou} C.-Y. Mou and P. B. Weichman, {\em Phys. Rev. E} {\bf 52},
3738 (1995).
\end{references}
\end{document}